\documentclass[aps,prl,nofootinbib,showpacs,twocolumn]{revtex4-1}
\usepackage{color}
\usepackage{bm}
\usepackage{graphicx}
\usepackage{amsmath}
\usepackage{amssymb}
\usepackage{enumitem}
\usepackage{hyperref}
\usepackage{subfigure}
\usepackage{array}
\usepackage[english]{babel}
\usepackage{tensor}
\usepackage{comment}
\usepackage[normalem]{ulem}
\usepackage[dvipsnames]{xcolor}

\def\be{\begin{equation}}
\def\ee{\end{equation}}
\def\ba{\begin{eqnarray}}
\def\ea{\end{eqnarray}}

\allowdisplaybreaks

\begin{document}

\title{CHAM: a fast algorithm of modelling non-linear matter power spectrum in the sCreened HAlo Model}

\author{Bin Hu$^{1}$} 
\email{bhu@bnu.edu.cn} 
\author{Xue-Wen Liu$^{2,3}$} 
\author{Rong-Gen Cai$^{2,3}$}
\affiliation{
$^{1}$Department of Astronomy, Beijing Normal University, Beijing, 100875, China\\
$^{2}$CAS Key Laboratory of Theoretical Physics, Institute of Theoretical Physics,\\
Chinese Academy of Sciences, P.O. Box 2735, Beijing 100190, China \\
$^{3}$School of Physical Sciences, University of Chinese Academy of Sciences, No. 19A Yuquan Road, Beijing 100049, China}

\begin{abstract}
We present a fast numerical screened halo model algorithm (\texttt{CHAM}) for modeling non-linear power spectrum for
the alternative models to $\Lambda$CDM. This method has three obvious advantages.
First of all, it is not being restricted to a specific dark energy/modified gravity model.
In principle, all of the screened scalar-tensor theories can be applied.
Second, the least assumptions are made in the calculation.
Hence, the physical picture is very easily understandable.
Third, it is very predictable and does not rely on the calibration from N-body simulation.
As an example, we show the case of Hu-Sawicki $f(R)$ gravity.
In this case, the typical CPU time with the current parallel Python script ($8$ threads) is roughly within $10$ minutes.
The resulting spectra are in a good agreement with N-body
data within a few percentage accuracy up to $k\sim1~h/{\rm Mpc}$.
\end{abstract}


\date{\today}

\maketitle



Several up-coming large scale structure surveys, such as Euclid~\footnote{\url{http://sci.esa.int/euclid}}, LSST~\footnote{\url{http://www.lsst.org}}, WFIRST~\footnote{\url{https://wfirst.gsfc.nasa.gov}}, are aiming to measure the matter power spectrum range  from $0.1$ to $10~{\rm Mpc}/h$ up to $1\%$ accuracy. One of the theoretical obstacles is how to model the non-linear power spectrum on these scales. By means of high precision simulation, we are able to
do this modelling with the requested accuracy within the $\Lambda$CDM paradigm. However, another scientific issue associated with  these activities, is to understand the nature of late-time cosmic acceleration, which is normally interpreted by dark energy (DE) or modified gravity (MG) models.

A number of alternative models to the $\Lambda$CDM  have inflated over the past decades.
The predictions on the background expansion, from most of the viable DE/MG models, are hard to be distinguished from the standard scenario.
Motivated by this fact, one of methods, is to focus on the parametrization of perturbation dynamics with given background cosmology.
For the linear perturbation, the effective field theory approach~\cite{Bloomfield:2012ff,Gubitosi:2012hu}, provides us a uniformed parametrization for scalar-tensor
type DE/MG models. The resulted linear Einstein-Boltzmann solvers, such as \texttt{EFTCAMB}~\cite{Hu:2013twa}\footnote{\url{http://www.eftcamb.org/}} are being
developed extensively.

As for the non-linear part, several different methods are being studied, such as N-body simulation~\cite{Winther:2015wla} and the hybrid scheme of the Lagrangian perturbation theory with N-body simulation~\cite{Valogiannis:2016ane}.
Take N-body simulation as an example, although this approach could provide an accurate result, it is too much expensive to construct the template for all the DE/MG models.
On the other hand, based on the current observations, the fifth force has to be shielded on the small scales, such as the solar system.
This results in the fact that, most of DE/MG models can be categorised into a few types, according to different screening mechanism, such as chameleon~\cite{Khoury:2003rn}, Vainshtein mechanism~\cite{Vainshtein:1972sx}, {\emph etc.}

This motives us to unify the non-linear spectrum modelling via the screening mechanism. Recently, Ref.~\cite{Lombriser:2016zfz} proposed a generic parametrization of the modified gravity.
In this parametrization, all the modified gravity effects are encoded into the scale-dependent gravitational constant
 \begin{eqnarray}
 \frac{G_{\rm eff}}{G_{\rm N}} &=& A + \sum_i^{N_0} B_{i} \prod_j^{N_i} b_{ij} \nonumber \\ &~&\cdot\left(\frac{r}{r_{0ij}}\right)^{a_{ij}}\left\{\left[1+\left(\frac{r_{0ij}}{r}\right)^{a_{ij}}\right]^{1/b_{ij}} - 1 \right\} \,, \label{eq:effscrchameleon}
\end{eqnarray}
where $A, B_i, r_{0ij}, a_{ij}, b_{ij}$ are the screening parameters.
With this great simplification, in this paper, we are aiming to present a fast and reasonably accurate algorithm for the non-linear spectrum in the DE/MG models.

\begin{figure}
\begin{center}
  \includegraphics[width=0.5\textwidth]{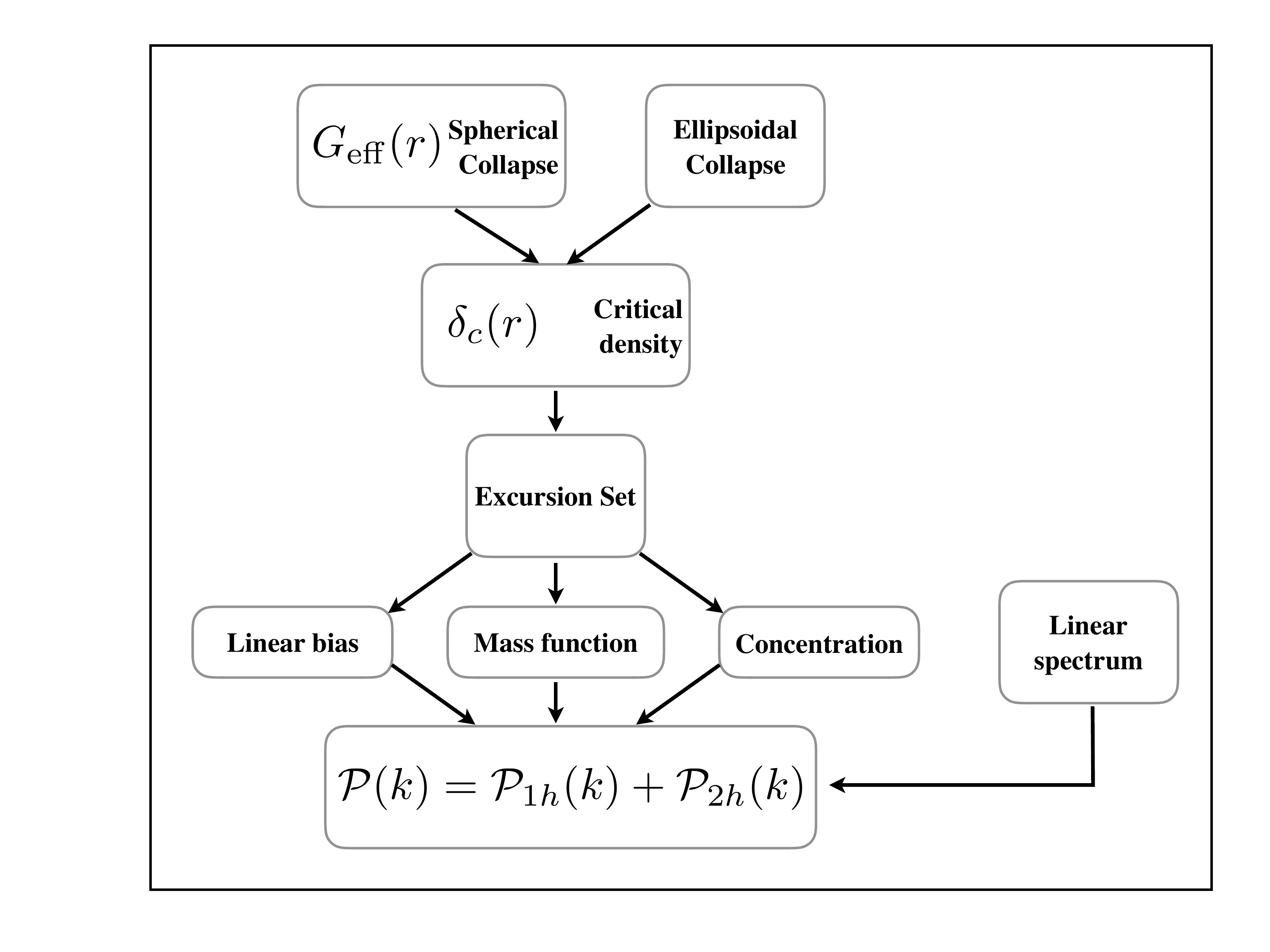}
  \caption{\label{fig:flow} Flow chart of sCreen HAlo Model (\texttt{CHAM}).}
\end{center}
\end{figure}

In details, our method is the extended halo model~\cite{Cooray:2002dia} with the scale-dependent gravitational constant given by Eq. (\ref{eq:effscrchameleon}).
The flow chart of our algorithm (namely, \texttt{CHAM}, denotes for sCreened HAlo Model) is presented in Fig.\ref{fig:flow}.
In the literature, the halo model for the DE/MG models has been extensively studied, such as \cite{Schmidt:2008tn,Lombriser:2013eza} for $f(R)$ gravity. 
Compared with them, our method has three advantages. First of all, this algorithm is not being restricted to a specific model.
Basically, all of the screened scalar-tensor theories can be applied. Secondly, the least assumptions are made in the calculation.
For example, instead of using the conventionally Sheth-Tormen mass function~\cite{Sheth:1999mn}, we solved the halo distribution function with a moving
barrier in terms of the excursion set formalism~\cite{Bond:1990iw,Zhang:2005ar}.
In particular, the scale dependence of critical density, $\delta_c$(r), is attributed not only to the MG effect in the process of spherical collapse, but
also to the ellipsoidal collapsing effect~\cite{Sheth:1999su}.
Last but not the least, our model is very predictive due to the fact that we have very limited parameters.


As a demonstration, we consider Hu-Sawicki $f(R)$ gravity model~\cite{Hu:2007nk}, which can satisfy the background $\Lambda$CDM expansion history
and evade the solar system tests:
\begin{eqnarray}
f(R)=-\bar{m}^2\frac{c_1(R/\bar{m}^2)^n}{c_2(R/\bar{m}^2)^n+1},~\bar{m}^2\equiv \kappa^2\bar{\rho}_{m0}/3,
	\label{eq:hsform}
\end{eqnarray}
where~$\bar m$ refers to the present Compton mass of the extra scalar field.
In the quasistatic but linear regime, the Poisson equation reads
\begin{eqnarray}
k^2 \Psi = -4 \pi G_{\rm N} \left( {4 \over 3} - {1 \over 3} {1 \over k^2\lambda_C^2 + 1} \right) a^2  \delta \rho_{\rm m}\,.
\label{eqn:nochameleon}
\end{eqnarray}
It is straightforward to see that below the Compton wavelength ($k\lambda_C\gg1$) of the extra scalar field, the gravitational constant
is enhanced by a factor $4/3$. Above this scale, the General Relativity is recovered.

In the non-linear high density regime, the fifth force carried by the scalar field is shielded by the chameleon mechanism.
Considering a spherical over density regime, this mechanism is very similar to the static electrodynamics phenomenon.
The scalar charge is only distributed on the surface, hence the thinner the surface is, the more significantly
the fifth force is screened
\begin{eqnarray}
 \frac{G_{\rm eff}}{G_{\rm N}} \approx 1 + \frac{1}{3}\min(3x-3x^2+x^3,1),~~x\equiv\frac{\Delta R}{R_{\rm TH}}\;,\nonumber\\
 \label{eq:Fenhancement}
\end{eqnarray}
where $x$ denotes the surface thickness, which can be parametrized schematically as~\cite{Lombriser:2016zfz}
\begin{eqnarray}
\frac{\Delta R}{R_{\rm TH}}\approx - C_1 r\left [(1+C_2 r^{-3})^{1/(\alpha-1)}-(1+C_3r^{-3})^{1/(\alpha-1)}\right],\nonumber\\
\end{eqnarray}
where $C_1,C_2,C_3$ and $\alpha$ can be read from the model parameters.
\begin{figure}
\begin{center}
  \includegraphics[width=0.5\textwidth]{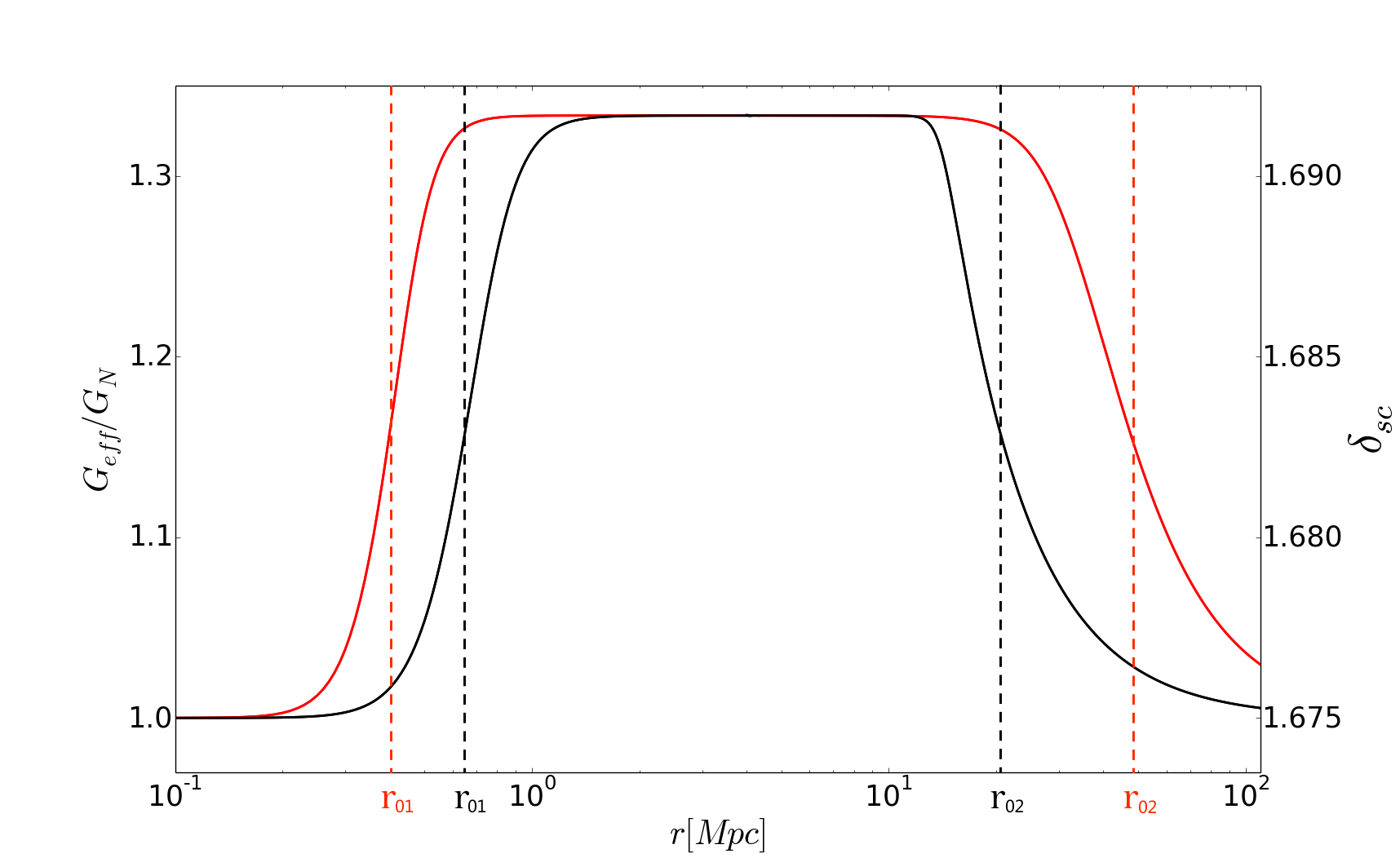}
  \caption{\label{fig:Geff} The scale dependent gravitational constant and critical density.
  The red and black curves represent for Hu-Sawicki model with $n=1$ and $f_{R0}=-10^{-4},-10^{-5}$, respectively.
  $r_{01},r_{02}$ label the averaged screening scale and Compton wavelength.}
\end{center}
\end{figure}
The above description can be concluded by Fig.\ref{fig:Geff}.
Below the averaged screening scale $(r_{01})$ the MG effect is shielded.
Above the Compton wavelength ($r_{02}$) the General Relativity is also recovered.
Between these two scales, the gravitational constant is enhanced by a factor $4/3$.


With the above model setup, the process of spherical collapse is accordingly modified by the scale dependent
gravitational constant. For simplicity, we firstly study the top-hat spherical collapse.
The Lagrangian radius can be solved in terms of  $y\equiv (r/r_i - a/a_i)$~\cite{Schmidt:2008tn}
\begin{eqnarray}
&& y''+\frac{H'}{H}y'=-\frac{1}{2}\frac{\Omega_ma^{-3}-2\Omega_{\Lambda}}{\Omega_ma^{-3}+\Omega_{\Lambda}}y
 -\frac{1}{2}\frac{\Omega_ma^{-3}}{\Omega_ma^{-3}+\Omega_{\Lambda}}\cdot  \nonumber\\
 && \frac{G_{\rm eff}(r)}{G_N}(\frac{a}{a_i}+y)\left[(\frac{1}{ya_i/a+1})^3(1+\delta_i)-1\right]\;,
  \label{eq:fulleqn}
\end{eqnarray}
where the prime is the derivative with respect to $\log a$ and $a_i,~r_i,~\delta_i$ denote the initial time, radius
and density of the corresponding Euclidean regime.

By the energy conservation law, as long as the initial (negative) potential energy dominates over the (positive) kinetic energy,
the initial over density patch, will always decouple from the background Hubble flow and finally collapse into a
virialized object. The collapsing threshold is proxied by the extrapolated linear density~$\delta_c$. In this work, we adjust
the initial value by asking the over density patch collapse at $a=1$. Hence, we integrate the following linear density
equation to the present time
 \begin{eqnarray}
 \delta''_m(r,a) + \left[2-\frac{3}{2}\Omega_m(a)\right]\delta'_m-\frac{3}{2}\frac{G_{\rm eff}(r)}{G_N}\cdot\Omega_m(a)\delta_m =0,\nonumber\\
 \label{eq:lineareqn}
\end{eqnarray}
with $G_{\rm eff}(r)/G_N$ described by Fig.\ref{fig:Geff}. 
We shall emphasize that unlike the $\Lambda$CDM case, the critical density in the modified gravity models is generally scale
dependent due to the scale dependent gravitational constant. This can also be seen from Fig.\ref{fig:Geff} (right vertical axis).
When the gravitational constant restores the Newtonian value, the spherical collapse critical density recovers $\delta_{sc}=1.676$.
When $G_{\rm eff}$ is enhanced, $\delta_{sc}$ reaches $1.692$.


Following the flow chart Fig.\ref{fig:flow}, we use the excursion set formalism~\cite{Press:1973iz,Bond:1990iw} to compute the probability $f(S)$ of forming a virialized object
with given linear matter fluctuation
\begin{eqnarray}
S(r)\equiv \int d^3k|\tilde{W}(kr)|^2P_L(k)\;.
\end{eqnarray}
At this step, let us consider a more realistic model, such as the ellipsoidal collapsing process. 
It will also introduce a scale dependent
critical density, such as~\cite{Sheth:1999su}
\begin{eqnarray}
\delta_c = \sqrt{a}\delta_{sc}[1+\beta (a\nu)^{-\alpha}],~~\nu \equiv \delta^2_{sc}/S,
\end{eqnarray}
where $\alpha=0.615,\beta=0.485$ and $\delta_{sc}$ is given by the spherical MG collapse.
Here we assumed that the effects of modified gravity and ellipsoidal collapse can be treated separately.
With the above moving barrier, the solution of distribution probability is given by \cite{Zhang:2005ar}
\begin{eqnarray}
f(S) &=& g(S) + \int^S_0 dS'f(S')h(S,S'),
\end{eqnarray}
in which
\begin{eqnarray}
g(S) &\equiv& \left[\frac{\delta_c}{S}-2\frac{d\delta_c}{dS}\right]P\left(\delta_c,S\right),\nonumber\\
h(S,S') &\equiv& \left[2\frac{d\delta_c}{dS}-\frac{\delta_c-\delta'_c}{S-S'}\right]P(\delta_c-\delta'_c,S-S'),
\label{eq:ghs}
\end{eqnarray}
where $P(\delta, S)$ is the normalized Gaussian distribution. This equation could be integrated numerically on an
mesh with equal spacing on: $S_i=i\Delta S$ with $i=0,1,\cdots,N$ and $\Delta S=S/N$.

\begin{figure}
\begin{center}
  \includegraphics[width=0.5\textwidth]{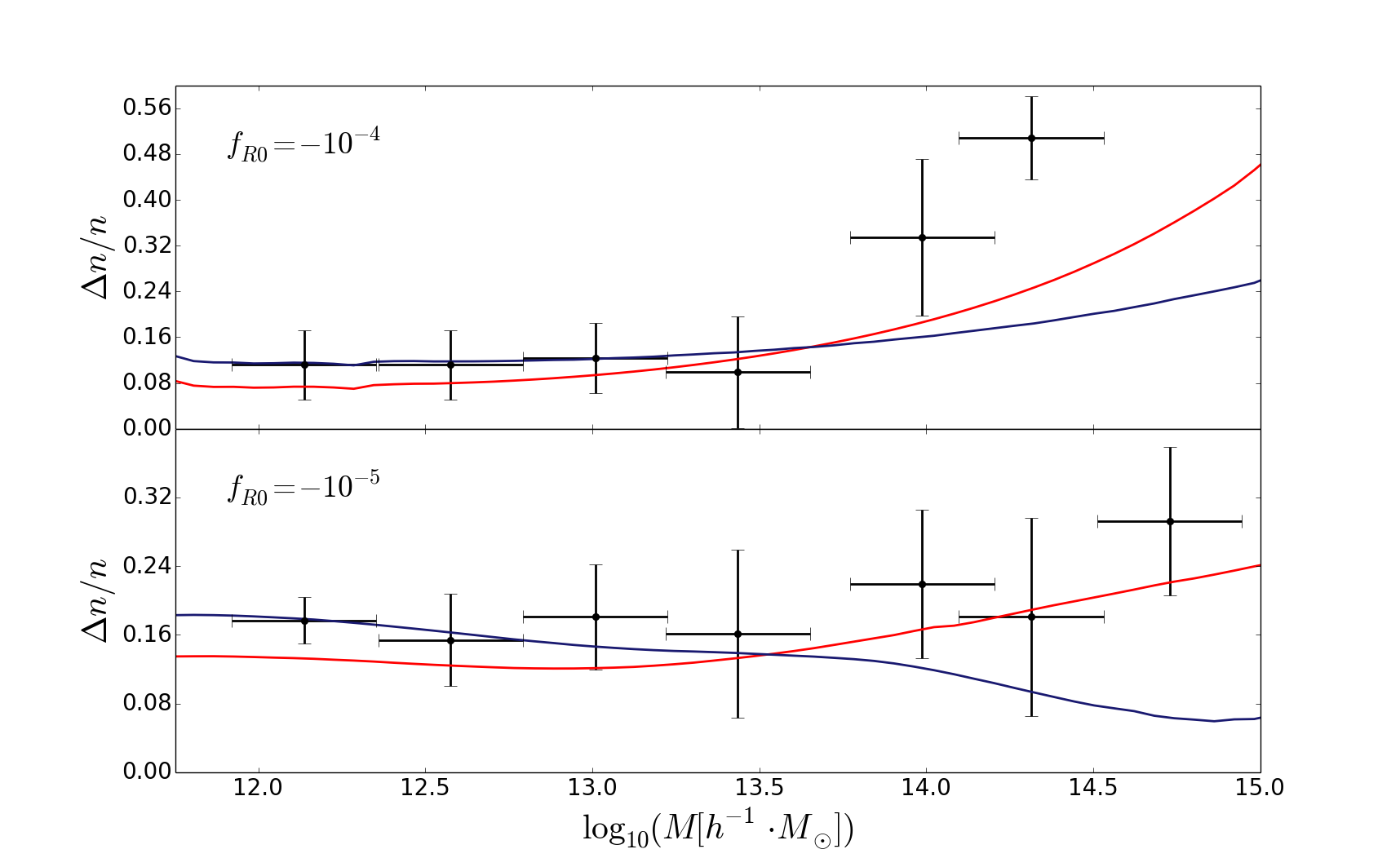}
  \caption{\label{fig:mass} Halo mass function comparison. Data points are from N-body simulation existed in the literature.
  The red and blue curves are those from Sheth-Tormen prescription and ours, respectively.}
\end{center}
\end{figure}


Armed with the distribution probability $f(\nu)$, we present the halo mass function in Fig.\ref{fig:mass},
where the data points are from N-body simulation in the literature~\cite{Li:2012by}.
We can see that \texttt{CHAM} prediction (blue) is systematically better than the Sheth-Tormen formalism (red),
in particular, in the low mass range. As for the high mass range, the two prescriptions behave statistically similar due to the
large scatters of the N-body data. 

The linear bias in \texttt{CHAM} is derived by using the peak-background split approach. In the large cell
limit, the mass inside a cell $M$ is much larger than the typical halo mass $m$, this leads to the number of haloes with mass $m$
inside the mass cell $M$ can be approximated by
\begin{eqnarray}
b_{\rm L}(M_{vir})&=& 1-\frac{\partial \ln n}{\partial \delta_{sc}}.
\end{eqnarray}
The derivative is realised by using Richardson four step interpolation to narrow the numerical error.

As for the density profile, we use the NFW form~\cite{Navarro:1995iw}
\begin{eqnarray}
 \rho(r) = \frac{\rho_s}{(r/r_s)(1+r/r_s)^2},
 \label{eqn:NFW}
\end{eqnarray}
where $r_s$ and $\rho_s$ are characteristic radius and density, which can be parametrized via the
concentration parameter
~\cite{Bullock:1999he}
\begin{eqnarray}
 c_{vir}(M_{vir}) & = & 9 \left[ \frac{M_*(M_{vir})}{M_{vir}} \right]^{0.13}, \label{eq:concentration}
\end{eqnarray}
where $M_{*}$ is defined via $\sigma(M_{*})=\delta_{sc}$. We shall emphasize again that the scale
dependence of $\delta_c$ leads to a different characteristic mass $M_*$ at different scales.

\begin{figure}
\begin{center}
  \includegraphics[width=0.5\textwidth]{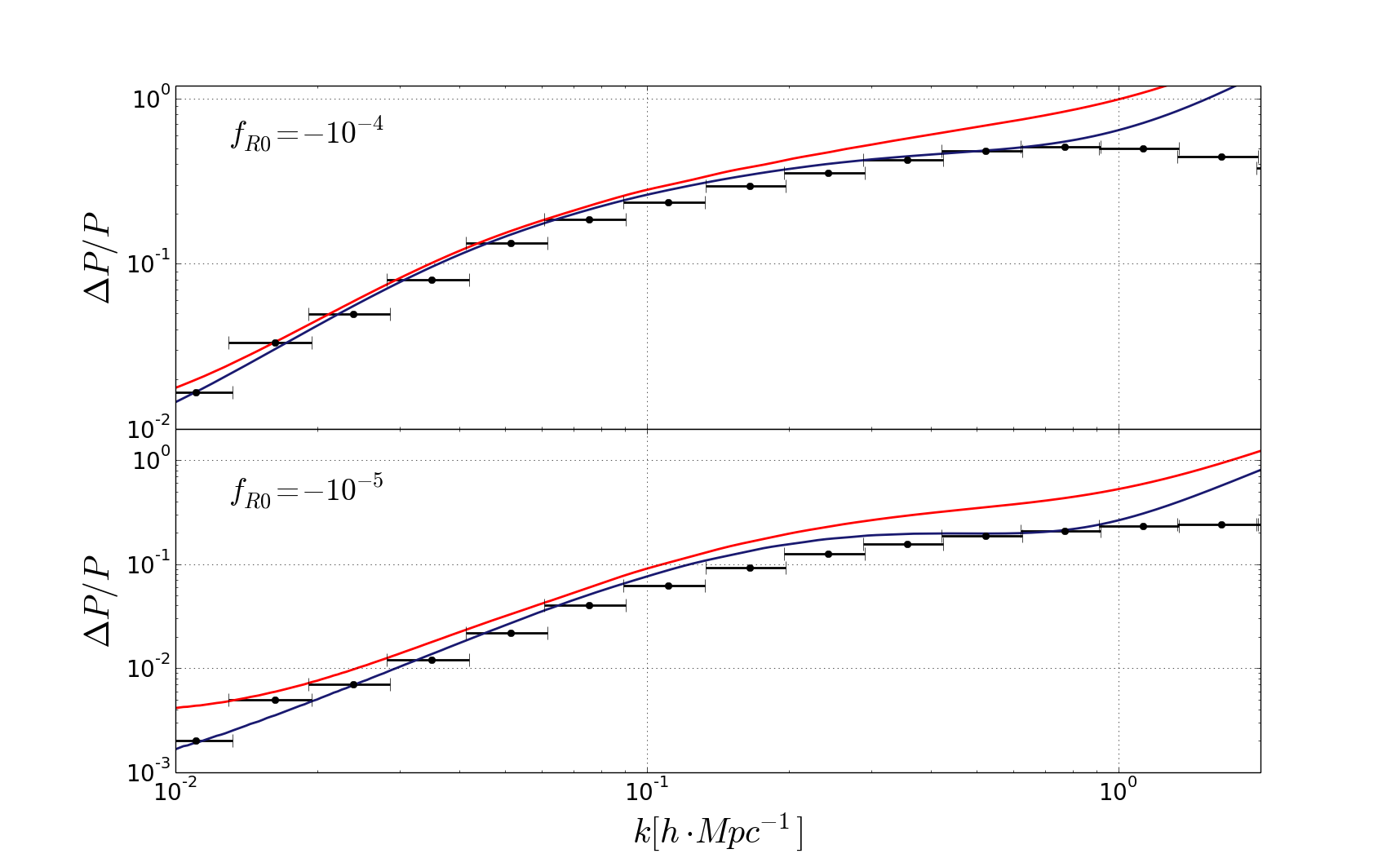}
  \caption{\label{fig:pok} Matter spectrum comparison. Data points are from N-body simulation  in the literature~\cite{Li:2012by}.
  The red and blue curves are those from Sheth-Tormen prescription and ours, respectively.}
\end{center}
\end{figure}

Now, we can assemble the mass function, linear bias as well as the concentration into the spectrum.
In the halo model, all mass is within individual halos. Hence, the correlation function is made out of two terms. One 
is the halo-halo correlation function which describes the density correlation between two halos on the large scale. 
The other is one halo term which describes the density correlation between two points inside one halo. 
This term shall dominate on the small scale. In Fourier space, the matter power
spectrum can be described as
\begin{eqnarray}
 P_{\rm mm}(k) & = & I^2(k) P_{\rm L}(k) + P^{1h}(k), \label{eq:nlpmm}
\end{eqnarray}
with
\begin{eqnarray}
 &&P^{1h}(k) =  \int d\ln M_{vir} n_{\ln M_{vir}} \frac{M_{vir}^2}{\bar{\rho}_m ^2} \left| y(k,M_{vir}) \right|^2 \label{eq:p1hk}\\
 &&I(k) = \int d \ln M_{vir} n_{\ln M_{vir}} \frac{M_{vir}}{\bar{\rho}_m} y(k,M_{vir}) b_{\rm L}, \label{eq:ik}
\end{eqnarray}
where $y(k,M)$ is the Fourier transform of NFW density profile and normalized as $\lim_{k\rightarrow0} y(k,M) = 1$.

Following this algorithm, we present our final spectrum at $z=0$ in Fig.\ref{fig:pok}.
The red and blue curves are Sheth-Tormen and \texttt{CHAM} prescriptions, respectively.
The same as the mass function results, in order to
verify our prediction, we compare the spectrum results with N-body simulation data from~\cite{Li:2012by}. In details,
we use the cosmology with ($\Omega_m=0.24,~\Omega_{\Lambda}=0.76,~H_0=73,~n_s=0.958,\sigma_8=0.8$).
The linear power spectrum is output from \texttt{EFTCAMB} Hu-Sawicki $f(R)$ module~\cite{Hu:2016zrh}.
From Fig.\ref{fig:pok}, we can clearly see that the \texttt{CHAM} spectrum results agree with N-body data within a
few percentage accuracy up to $k\sim1~h/{\rm Mpc}$.
We shall emphasize here that, unlike the halofit philosophy, in this model, \texttt{CHAM} only has two parameters, namely
$r_{01}, r_{02}$, which denote  the averaged screening and Compton wavelength scales, respectively.

In conclusion, we developed a fast numerical halo model algorithm for modelling non-linear power spectrum for
the alternative models to $\Lambda$CDM. As an example, we show the case of Hu-Sawicki $f(R)$ gravity. The typical
CPU time with the current parallel Python script\footnote{\texttt{CHAM} repository: \url{https://github.com/hubinitp/CHAM}} ($8$ threads) is roughly within $10$ minutes.  The resulting spectra are in a good agreement with N-body
data within a few percentage accuracy up to $k\sim1~h/{\rm Mpc}$. More importantly, this method is very predictive
and it does not ask for the calibration from N-body simulation. 
We believe this method can be widely used in several aspects of data analysis, such as covariance matrix, parameter estimation, {\it etc}.

\vspace{0.1cm}


\begin{acknowledgments}
We thank Hou-Jun Mo, Zu-Hui Fan, Lucas Lombriser for helpful discussion.
BH is supported by the Beijing Normal University Grant under the reference No. 312232102,
Chinese National Youth Thousand Talents Program and the Fundamental Research Funds for
the Central Universities under the reference No. 310421107. XWL and RGC are supported by  the National Natural Science Foundation of China Grants No.11690022,  No.11435006 and No.11647601, and by the Strategic Priority Research Program of CAS Grant No.XDB23030100 and by the Key Research Program of Frontier Sciences of CAS.
\end{acknowledgments}

\appendix


\bibliographystyle{apsrev4-1}

\bibliography{cham}

\end{document}